\documentclass[prd,showpacs,showkeys,twocolumn]{revtex4}
\usepackage{bm}
\input{epsf}
\begin{document}
\title{R=0 spacetimes and self-dual Lorentzian wormholes}
\author{Naresh Dadhich}
\email{nkd@iucaa.ernet.in}
\affiliation{Inter--University Centre for Astronomy and Astrophysics \\
Post Bag 4, Ganeshkhind, Pune 411 007, India}
\author{Sayan Kar}
\email{sayan@cts.iitkgp.ernet.in}
\affiliation{Department of Physics and Centre for Theoretical Studies
\\
Indian Institute of Technology, Kharagpur 721 302, WB, India}
\author{Sailajananda Mukherjee}
\email{sailom@nbu.ernet.in}
\affiliation{Department of Physics, University of North Bengal
\\
P.O. North Bengal University, Siliguri 734430 Dist Darjeeling, WB, India}
\author{Matt Visser}
\email{visser@kiwi.wustl.edu}
\homepage{http://www.physics.wustl.edu/~visser}
\affiliation{Physics Department, Washington University,
Saint Louis MO 63130--4899, USA}
\author{\ } 
\begin{abstract}
A two--parameter family of spherically symmetric, static Lorentzian
wormholes is obtained as the general solution of the equation
$\rho=\rho_t=0$, where $\rho = T_{ij}\;u^iu^j$, $\rho_t = (T_{ij} -
{1\over2} T\,g_{ij})\;u^iu^j$, and $u^i\;u_i =- 1$. This equation
characterizes a class of spacetimes which are ``self dual'' (in the
sense of electrogravity duality). The class includes the Schwarzschild
black hole, a family of naked singularities, and a disjoint family of
Lorentzian wormholes, all of which have vanishing scalar curvature
($R=0$).  Properties of these spacetimes are discussed. Using
isotropic coordinates we delineate clearly the domains of parameter
space for which wormholes, nakedly singular spacetimes and the
Schwarzschild black hole can be obtained.  A model for the required
``exotic'' stress--energy is discussed, and the notion of
traversability for the wormholes is also examined.
\end{abstract}
\maketitle
\def\d{{\mathrm{d}}}
\section{Introduction}
Traversable Lorentzian wormholes have been in vogue ever since Morris,
Thorne and Yurtsever {\cite{mty:prl88}} came up with the exciting
possibility of constructing time machine models with these exotic
objects. The seminal paper by Morris and Thorne {\cite{mt:ajp88}}
demonstrated that the matter required to support such spacetimes
necessarily violates the null energy condition. This at first made
people rather pessimistic about their existence in the classical
world. Semiclassical calculations based on techniques of quantum
fields in curved spacetime, as well as an old theorem due to Epstein,
Glaser and Yaffe {\cite{egy:inc}}, raised hopes about the generation
of such spacetimes through quantum stresses. The Casimir effect was
put to use by MTY themselves to justify their introduction of such
spacetimes as a means of constructing model time machines.

In the last twelve years or so there have been innumerable attempts at
solving the so-called ``exotic matter problem'' in wormhole physics.
(For a detailed account of wormhole physics up to 1995 see the
book by Visser {\cite{visser:book}}. For a slightly later survey
see~\cite{geometric}.)  Alternative theories of gravity
{\cite{alt:prd}}, evolving (dynamic, time--dependent) wormhole
spacetimes {\cite{sk:prd,mv:prd,hochberg}} with varying definitions of
the throat have been tried out as possible avenues of resolution.

Despite multiple efforts, all these spacetimes still remain in the
domain of fiction. At times, their simplicity makes us believe that
they just might exist in nature though we are very far from actually
seeing them in the real world of astrophysics. (For attempts towards
astrophysical consequences see {\cite{ms:01}}.)

This paper does not set out to solve the exotic matter problem.  On
the contrary, it does have exotic matter as the source once again. The
obvious query would therefore be --- what is actually new?  The first
novelty is related to the method of construction. Generally, a
wormhole is constructed by imposing the geometrical requirement on
spacetime that there exists a throat but no horizon. This is however
not couched in terms of an equation restricting the stress-energy.

Perhaps for the first time, we are proposing a specific restriction on
the form of the stress-energy that, when solving the Einstein
equations, automatically leads to a class of wormhole solutions.  The
characterization we have in mind for our class of static self-dual
wormholes is
\begin{equation}
\rho = \rho_t = 0,
\label{E:basic}
\end{equation}
where $\rho$ and $\rho_t$ are respectively the energy density measured
by a static observer and the convergence density felt by a timelike
congruence. (Applying both of these conditions, plus the Einstein
equations, implies $R=0$). It is remarkable that the general solution
{\cite{nkd:cs}} of this equation automatically incorporates the basic
characteristics of a Lorentzian wormhole. A generic wormhole spacetime
is an {\emph{ad hoc}} construction and hence no equation would be able
to encompass all wormhole spacetimes. Equation (\ref{E:basic})
indicates that space is empty relative to timelike particles as they
encounter neither energy density nor geodesic convergence.  This, in
turn, leads to the fact that the general spacetime will be a
modification of the Schwarzschild solution, which would be contained
in the general solution of equation (\ref{E:basic}) as a special
case.  Portions of the general solution space are interpreted as
wormhole spacetimes with the throat given by the well known
Schwarzschild radius, which no longer defines the horizon.
Indeed equation (\ref{E:basic}) uniquely characterizes a class of
``self dual'' static wormhole spacetimes which contains the
Schwarzschild solution.  The notion of duality we have in mind
involves the interchange of the active and passive electric 
parts of
the Riemann tensor (termed as {\em electrogravity duality}). This
duality, which leaves the vacuum Einstein equation invariant, was
defined by one of the present authors in {\cite{nkd:98}}.
Electrogravity duality essentially implies the interchange of the
Ricci and Einstein tensors.  For vanishing Ricci scalar, these two
tensors become equal and the corresponding solutions could therefore
be called {\em self-dual} in this sense. Under the duality
transformation, $\rho$ and $\rho_t$ are interchanged indicating
invariance of equation (\ref{E:basic}).  Since energy densities vanish
and yet the spacetime is not entirely empty, the matter distribution
would naturally have to be exotic (violating all the energy conditions
{\cite{visser:book}}).  Physically, the existence of such spacetimes
might be doubted because of this violation of the weak and null energy
conditions.  Analogous to the spatial--Schwarzschild wormhole, for
which $g_{00}=-1$ and $g_{11}=({1-\frac{2m}{r}})^{-1}$, these
spacetimes have zero energy density but nonzero pressures. The
spatial--Schwarzschild wormhole is one specific particular solution of
the equations $\rho = 0$, $\rho_t = 0$, while here we exhibit the most
general solution {\cite{nkd:cs}}.

Furthermore, on rewriting the line element in terms of isotropic
coordinates we realise the existence of all three classes of
spherically symmetric spacetimes --- namely the black holes, the
wormholes and the nakedly singular geometries --- all within the
framework of our general solution.  This happens through the
understanding of the nature of the spacetime for different domains of
the two parameters $\kappa$ and $\lambda$ defined below. A detailed
discussion on this is included too.

The following section (\ref{S:R=0}) deals with all the above mentioned
aspects of $R=0$ spacetimes.  In the last section
(\ref{S:conclusions}) we offer our conclusions and remarks.  We choose
the metric signature as $(- + + +)$ and set $c=1$ unless otherwise
stated.

\section{$R=0$ characterization of Lorentzian wormholes}
\label{S:R=0}
The Lorentzian wormhole, {\em a la} Morris and Thorne, is defined
through the specification of two arbitrary functions $b(r)$ and
$\phi(r)$ which appear in the following generic version of a
spherically symmetric, static line element:
\begin{equation}
ds^2 = -e^{2\phi (r)}\d t^2 + \frac{\d r^2}{1-\frac{b(r)}{r}}
+ r^2\left (\d\theta^2 + \sin^2 \theta \;\d\varphi^2 \right ).
\end{equation}
The properties of $b(r)$ and $\phi(r)$ which `make' a wormhole
are~\cite{mt:ajp88,visser:book}:
\begin{enumerate}
\item[(a)]
A {\em no-horizon} condition $\rightarrow$ $ e^{2\phi}$ has no zeros.
\\
(The function $\phi(r)$ is called the red--shift function.)
\item[(b)]
A wormhole shape condition $\rightarrow$ \quad $b(r=b_0)=b_0$; \hfill
\\
with $\frac{b(r)}{r} \leq 1$ ($\forall r\geq b_0$).
\\
($b(r)$ is called the wormhole shape function.)
\item[(c)]
Asymptotic flatness:
$\frac{b(r)}{r} \rightarrow 0$ as  $r\rightarrow \infty$.
\end{enumerate}
These features provide a minimum set of conditions which lead, through
an analysis of the embedding of the spacelike slice in a Euclidean
space, to a geometry featuring two asymptotic regions connected by a
bridge. Topologically different configurations where we only have one
asymptotic region was the origin of the name `wormhole' {\em a la}
Wheeler.  It is well--known~\cite{mt:ajp88,visser:book} that these
conditions lead to the requirement that the stress-energy which
supports the wormhole violates the null energy condition (and even the
averaged null energy condition)~\cite{mv:prd,hochberg}.

Amongst examples, the simplest is of course the spatial-Schwarzschild
wormhole defined by the choice $\phi = 0$ and $b(r)=2m$. The horizon
is (by fiat) gotten rid of simply by choosing $g_{00}=-1$ and the
wormhole shape is retained by choosing $b(r)=2m$. This geometry would
of course be contained in the general solution of equation
(\ref{E:basic}). Many other examples can be constructed. There is no
general principle as such to generate these wormholes --- one might
just `tailor-make' them according to ones taste. If we demand that the
wormhole spacetime must contain the Schwarzschild spacetime, the
equation (\ref{E:basic}) completely and uniquely characterizes it.
\subsection{General solution}
\label{SS:general-solution}
In order to make a wormhole, we have to specify/determine two
functions.  Generally one of them is chosen {\emph{by fiat}} while the
other is determined by implementing some physical condition. In this
paper, we have proposed the equation (\ref{E:basic}) as the equation
for wormhole, which would imply $R=0$, and place a constraint on the
wormhole shape function. Alternatively, we could choose the shape
function and solve for $R=0$. The $R=0$ constraint will be a condition
on $b(r)$ and $\phi(r)$ and its derivatives. In an earlier paper
{\cite{skds:prd}} $\phi(r)$ was chosen and an appropriate $b(r)$
(satisfying wormhole conditions) was obtained as a solution to the
$R=0$ constraint. Here, we do the opposite, first demand $\rho=0$ and
then solve for $R=0$, which would determine both $\phi$ and
$b$. Interestingly, the most general solution of equation
(\ref{E:basic}) automatically incorporates the requirement of
existence of a throat without horizon. This is thus a natural
characterization of a Lorentzian wormhole.

Defining the diagonal energy momentum tensor components as $T_{00} =
\rho (r)$, $T_{11}=\tau (r)$ and $T_{22}=T_{33}=p(r)$ and using the
Einstein equations with the assumption of the line element given above
we find that:
\begin{widetext}
\begin{eqnarray}
\rho (r) &=& \frac{1}{8\pi\; G} \;
\frac{b'}{r^2};
\\
\tau (r) &=&  \frac{1}{8\pi\; G} \;
\left[
- \frac{b}{r^3} + 2 \frac{\phi'}{r} \left (1-\frac{b}{r}\right )
\right];
\\
p(r) &=& \frac{1}{8\pi\; G} \;
\left\{
\left( 1-\frac{b(r)}{r} \right)
\left [\phi '' -\frac{b'r-b}{2r(r-b)}
\phi ' + {\phi'}^2 + \frac{\phi '}{r} - \frac{b'r-b}{2r^2(r-b)} \right]
\right\}.
\end{eqnarray}
\end{widetext}
(Note that $\tau$ as defined above is simply the radial pressure
$p_r$, and differs by a minus sign from the conventions
in~\cite{mt:ajp88,visser:book}. The symbol $p$ is simply the transverse
pressure $p_t$.)  Implementing the condition $\rho = \rho_t$ (or,
equivalently $ R = T = 0$) we find the following equation:
\begin{equation}
\xi' + \xi^{2} + \left (\frac{2}{r}-\frac{b'r-b}{2r(r-b)} \right ) \xi
= \frac{b'}{r(r-b)},
\label{E:riccati}
\end{equation}
where $\xi = \phi'$ and the prime denotes a derivative with respect to
$r$.  Given $b(r)$ we can solve the above equation to obtain
$\phi$. We note that the above equation (with a given $b$) is a nonlinear,
first-order ordinary differential equation which is known in
mathematics as a Riccati equation. This equation is covariant under
fractional linear transformations of the dependent variable [SL(2,R)
symmetry].  In principle equation (\ref{E:riccati}) can be thought of
as the `master' differential equation for all spherically symmetric,
static $R=0$ spacetimes, examples of which include the Schwarzschild
and the Reissner--Nordstrom geometries.

Now solving for $\rho = 0$ clearly gives $b = \mathrm{constant} =
2m$. Then equation (\ref{E:riccati}) simplifies to:
\begin{equation}
\xi'+\xi^{2} + \frac{\xi}{r}\left ( \frac{2r-3m}{r-2m}\right ) =0.
\label{E:riccati2}
\end{equation}
It admits the most general solution given by
\begin{equation}
g_{00} = - \left (\kappa + \lambda \sqrt{1-\frac{2m}{r}}\right )^2,
\end{equation}
where $\kappa$ and $\lambda$ are constants of integration.

Clearly the Schwarzschild geometry is the special solution for which
$R_{ij}=0$, to which the general solution reduces when $\kappa=0$.
This shows that the Schwarzschild solution is contained in the above
general solution. It also contains the spatial-Schwarzschild wormhole
with $g_{00} = -1$ when $\lambda = 0$.  The solution then admits no
horizon but there is a wormhole throat at $r = 2m$. We thus have a
Lorentzian wormhole.

The components of the energy momentum tensor for this geometry turn
out to be:
\begin{eqnarray}
\rho &=& 0;
\\
\tau &=& - \frac{1}{8\pi\; G}
\left [
\frac{2m\kappa}{r^3 \left(\kappa + \lambda\sqrt{1-\frac{2m}{r}}\right)}
\right ];
\\
p &=& \frac{1}{8\pi\; G}
\left [
\frac{m\kappa}{r^3 \left(\kappa + \lambda\sqrt{1-\frac{2m}{r}}\right)}
\right ].
\end{eqnarray}
The weak ($\rho \ge 0$, $\rho + \tau \ge 0$, $\rho + p \ge 0$) and
null ($\rho + \tau \ge 0$, $\rho + p\ge 0$) energy conditions are both
violated.  We note that the stress--energy given above satisfies
$\rho+\tau = - 2p$; which obviously follows from $R=0 \quad
\Rightarrow \quad T=0$.  The violation of the energy condition stems
from the violation of the inequality $\rho+\tau \ge 0$. The extent of
the violation, caused by the $1/{r^3}$ behaviour of the relevant
quantity, is large in the vicinity of the throat.  One does have a
control parameter $\kappa$, which, can be chosen to be very small
in order to restrict the amount of violation.

\begin{widetext}
The complete line element for the geometry discussed above is:
\begin{equation}
\d s^2 =
- \left(\kappa+\lambda\sqrt{1-{2m\over r}}\right)^2 \;\d t^2
+ {\d r^2\over1-2m/r}
+ r^2 \left(\d\theta^2 + \sin^2\theta \;\d\varphi^2 \right).
\label{E:gen1}
\end{equation}
Of course we could also consider the line element below (obtained by
replacing $\lambda$ by $-\lambda$), which also has $R=0$.
\begin{equation}
\d s^2 =
- \left(\kappa-\lambda\sqrt{1-{2m\over r}}\right)^2  \;\d t^2
+ {\d r^2\over1-2m/r}
+ r^2 \left(\d\theta^2 + \sin^2\theta \;\d\varphi^2 \right).
\label{E:gen2}
\end{equation}
%
Note that these metrics only make sense (by construction) for
$r\geq2m$. So to really make the wormhole explicit we need two
coordinate patches, $r_1\in(2m,\infty)$, and $r_2\in(2m,\infty)$,
which we then have to sew together at $r=2m$. (And in this particular
case the geometry is smooth across the junction provided we pick the
$+$ root on one side and the $-$ root on the other side, in which case
there is no thin shell contribution at the junction.)  This is not
particularly obvious, and to make this a little clearer, it is
convenient to go to isotropic coordinates, defined by
\begin{equation}
r = \bar r \left(1+{m\over2\bar r}\right)^2 .
\end{equation}
Since the space part of the metric for the general solution
[(\ref{E:gen1}) or (\ref{E:gen2})] is identical to the space part of
the Schwarzschild geometry we can use exactly the same transformation
for going from curvature coordinates to isotropic coordinates as was
used for Schwarzschild itself.  Then it is easy to see that
%
\begin{equation}
\d s^2 =
- \left\{
\kappa+\lambda\left[{1-m/2\bar r\over1+m/2\bar r}\right]
\right\}^2  \;\d t^2
+ \left(1+{m\over2\bar r}\right)^4
 \left[\d \bar r^2
 + \bar r^2 \left(\d\theta^2 + \sin^2\theta \;\d\varphi^2 \right)
\right]
\label{E:general}
\end{equation}
\end{widetext}
The space part of the geometry is invariant under inversion
$\bar r\to m^2/(4\bar r)$.

The advantage of isotropic coordinates is that in almost all cases a
single coordinate patch covers the entire geometry, $\bar r\approx 0$
is the second asymptotically flat region. Indeed whenever the geometry
is such that it can be interpreted as a Lorentzian wormhole then the
isotropic coordinate patch is a global coordinate patch.  (This is not
a general result; it works for the class of geometries in our general
solution [(\ref{E:gen1}), (\ref{E:gen2}), or (\ref{E:general})]
because the space part of the metric is identical to Schwarzschild.)

We now have a single global coordinate patch for the (alleged)
traversable wormhole, and use it to discuss the the properties of the
geometry (we always take $m>0$ since otherwise there is an
unavoidable naked singularity in the space part of the metric,
regardless of the values of $\lambda$ and $\kappa$):

\begin{itemize}
\item
1) The geometry is invariant under simultaneous sign flip
$\lambda\to-\lambda$, $\kappa\to-\kappa$; it is also invariant under
simultaneous inversion $\bar r\to m^2/(4\bar r)$ and sign reversal
$\lambda\to-\lambda$ (keeping $\kappa$ fixed).
\item
2) $\kappa=0$, $\lambda\neq0$ is the Schwarzschild geometry; it is
non-traversable.
\\
(This is also an example of a case where even the isotropic coordinate
system does not cover the entire manifold.)
\item
3) $\lambda=0$, $\kappa\neq0$ is the spatial-Schwarzschild traversable
wormhole.
\\
(And here clearly the isotropic coordinate system does cover the
entire manifold.)
\item
4) $\lambda=0$, $\kappa=0$ is singular.
\item
5) At the throat $g_{tt}(r=2m) = -\kappa^2$, so
$\kappa\neq0$ is required to ensure traversability.
\item
6) Is there ever a ``horizon''? This requires a little analysis.
\end{itemize}
A horizon would seem to form if $g_{tt}$ has a zero, that is if there
is a physically valid solution to
\begin{equation}
\kappa(1+m/2r)+\lambda(1-m/2r)=0.
\end{equation}
Solving this equation we obtain
\begin{equation}
r_H = {m\over2} {\lambda-\kappa\over\lambda+\kappa}.
\end{equation}
That is, a horizon tries to form (though typically not at the throat)
if
\begin{equation}
{\lambda-\kappa\over\lambda+\kappa} > 0.
\end{equation}
Unfortunately, this ``would be horizon'' is actually a naked
singularity. To see this we calculate
\begin{equation}
\tau =
-{128\over8\pi\;G} \; {\kappa m r^3
\over
(2r+m)^5 \; (2[\kappa+\lambda]r + [\kappa-\lambda] m) };
\end{equation}
\begin{equation}
p =
+{64\over8\pi\;G} \; {\kappa m r^3
\over
(2r+m)^5 \; (2[\kappa+\lambda]r + [\kappa-\lambda] m) };
\end{equation}
and notice that the radial and transverse pressure both diverge as
$g_{tt}\to0$. In fact it is better to rewrite the above as
\begin{equation}
\tau =
-{128\over8\pi\;G} \; {\kappa m r^3
\over
(2r+m)^6 \; \sqrt{-g_{tt} }};
\end{equation}
\begin{equation}
p =
+{64\over8\pi\;G} \; {\kappa m r^3
\over
(2r+m)^6 \; \sqrt{-g_{tt} }};
\end{equation}
explicitly showing that $g_{tt}\to0$ is a naked curvature singularity.
To reiterate, this curvature singularity forms if
\begin{equation}
{\lambda-\kappa\over\lambda+\kappa} > 0.
\end{equation}
This occurs if either
\begin{equation}
\lambda+\kappa >0 \qquad \hbox{and} \qquad \lambda-\kappa >0;
\end{equation}
or
\begin{equation}
\lambda+\kappa <0 \qquad \hbox{and} \qquad \lambda-\kappa <0.
\end{equation}
Outside of these regions the curvature singularity does not form, the
$g_{tt}$ component of the metric never goes to zero, and we have a
traversable wormhole.

\begin{figure}

\centerline{\epsfxsize=3.25in\epsffile{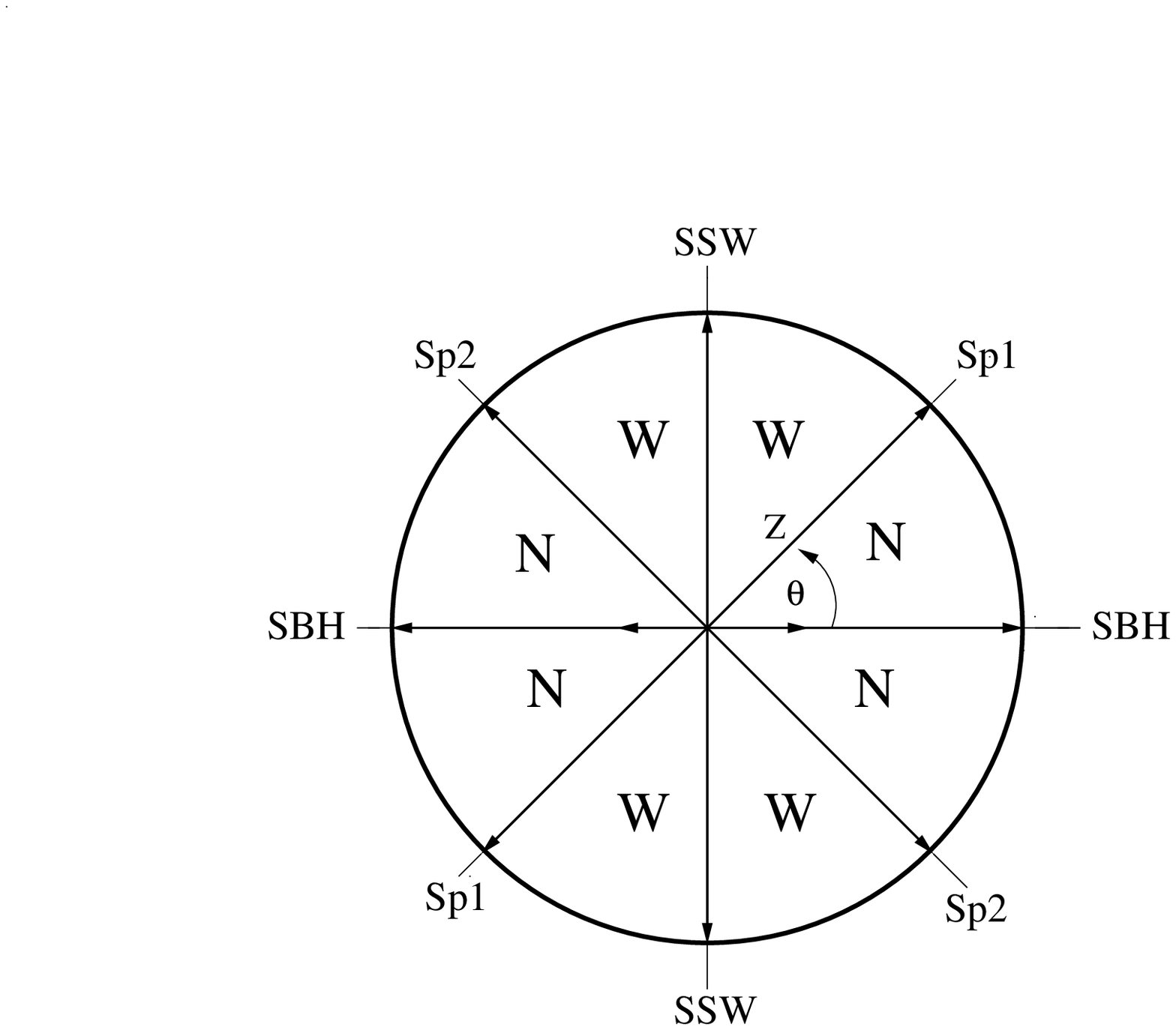}}


\caption{The solution space showing different regions
representing naked singularities, wormholes and black holes. Here
$\kappa= Z\sin \theta$ and $\lambda = Z\cos \theta$. N--Naked
singularities, W--Wormholes, SBH--Schwarzschild black hole,
SSW--Spatial Schwarzschild Wormhole, Sp1,2--Special solutions.}

\end{figure}

To summarize: The $\kappa$--$\lambda$ plane has the following
structure (let $\kappa$ run up the vertical axis; and define $\lambda
= Z\;\cos\theta$; $\kappa= Z\;\sin\theta$, see figure 1):
\begin{itemize}
\item
$\theta=0$ (the $+\lambda$ axis): Schwarzschild spacetime.
\item
$\theta\in(0,\pi/4)$ naked singularity.
\item
$\theta=\pi/4$ special; see below.
\item
$\theta\in(\pi/4,3\pi/4)$ traversable wormhole;
\\
$\theta=\pi/2$ (the $+\kappa$ axis) is the spatial-Schwarzschild wormhole.
\item
$\theta=3\pi/4$ special; see below.
\item
$\theta\in(3\pi/4,\pi)$ naked singularity.
\item
$\theta=\pi$ ($-\lambda$ axis): Schwarzschild spacetime.
\item
$\theta>\pi$: repeat the previous diagram in the lower half plane.
\end{itemize}
Let us now look at the two special cases:
\begin{itemize}
\item
$\theta=\pi/4 \quad \Rightarrow \quad \lambda=\kappa$. The geometry is:
\begin{eqnarray}
\d s^2 &=&
- \left\{{2\kappa\over1+m/2\bar r}\right\}^2  \;\d t^2
\\
&&+ \left(1+{m\over2\bar r}\right)^4
 \left[\d \bar r^2
 + \bar r^2 \left(\d\theta^2 + \sin^2\theta \;\d\varphi^2 \right)
\right].
\nonumber
\end{eqnarray}
The $\bar r \rightarrow 0$ region is not flat. (Space is
asymptotically flat, but spacetime isn't since $g_{tt}\to 0$ as $\bar r\to
0$.)
\item
$\theta=3\pi/4 \quad \Rightarrow \quad \lambda=-\kappa$. The geometry
is:
\begin{eqnarray}
\d s^2 &=&
- \left\{{\kappa m/{\bar r}\over1+m/2\bar r}\right\}^2  \;\d t^2
\\
&&
+ \left(1+{m\over2\bar r}\right)^4
 \left[\d \bar r^2
 + \bar r^2 \left(\d\theta^2 + \sin^2\theta \;\d\varphi^2 \right)
\right].
\nonumber
\end{eqnarray}
The $\bar r\rightarrow\infty$ region is not flat. (Space is
asymptotically flat, but spacetime isn't since $g_{tt}\to 0$ as $\bar r\to
\infty$.)
\end{itemize}
Other interesting features are:
\begin{equation}
g_{tt}(r=\infty) = -(\kappa+\lambda)^2
\end{equation}
\begin{equation}
g_{tt}(r=0) = -(\kappa-\lambda)^2
\end{equation}
That is, time runs at different rates in the two asymptotic
regions. If we try to reconnect the other side of the wormhole back to
our own universe we get a ``locally static'' wormhole in the sense of
Frolov~\cite{frolov} and Novikov~\cite{novikov}.
\subsection{Matter fields}
\label{SS:matter-fields}
We now move on briefly to the generation of the stress energy for the
two--parameter asymptotically flat wormhole in $3+1$ dimensions.  (The
two parameters are $m$ and $\kappa$.)
As is
well--known, the energy--momentum tensor which acts as a source for
the Schwarzschild spacetime with a global monopole can be generated
through a triplet of scalar fields $\phi^{a}$ self-interacting via a
Higgs potential~\cite{bv:gm}. The Lagrangian for this is:
\begin{equation}
L_{\mathrm{scalar}} =
\frac{1}{2} \partial_{\mu}\phi^{a}\;\partial^{\mu}\phi^{a}
+\frac{1}{4}\zeta \left ( \phi^a \phi^a-\eta^2\right )^{2}
\end{equation}
Choosing a monopole-like field configuration $\phi^{a} = \eta \;f(r)
\;{x^a}/{r}$ we can generate the required stress energy in the
region away from the core of the monopole (where $f(r) \equiv 1$ ).
Motivated by this model with a triplet of scalar fields, we make an
attempt to generate only $\tau$ and $p$ without making any
contribution to $\rho$ as required by the Einstein tensor for the line
element of the general wormhole discussed in the previous subsection
(\ref{SS:general-solution}).  To this end, we introduce a Lagrangian
with a triplet $\psi^{a}$ given by
\begin{equation}
L_{\mathrm{scalar}} =
\frac{1}{2}\partial_{\mu}\psi^{a}\;\partial^{\mu} \psi^{a} + \frac{1}{6}\sigma
\left (\psi^{a}{\psi^a}\right )^{3}
\end{equation}
Assuming a general metric of the form:
\begin{equation}
\d s^2 = - B(r) \;\d t^2 + A(r)\; \d r^2 +
r^2 \left(\d\theta^2 + \sin^{2}\theta\;\d\varphi^2 \right),
\end{equation}
and $\psi^{a}=g(r)\;{x^a}/{r}$ we obtain the differential equation
for the function $g(r)$:
\begin{equation}
\frac{1}{A} g'' +
\left[ \frac{2}{Ar} + \frac{1}{2B} \left(\frac{B}{A}\right)'\right] g'
-\frac{2g}{r^2} - \sigma g^5 = 0.
\end{equation}
With $g(r)={a}/{\sqrt{r}}$ the above equation results in a constraint
on the parameters $a$ and $\sigma$:  $\sigma a^4 = -{9}/{4}$, to
leading order in ${1}/{r}$.

The components of the energy momentum tensor are:
\begin{eqnarray}
\rho &=&
T_{00} =
\frac{1}{8\pi\; G}
\left [\frac{g'^{2}}{2A} + \frac{g^2}{r^2} + \frac{\sigma}{6}g^6 \right ];
\\
\tau &=& T_{11} =
\frac{1}{8\pi\; G}
\left [\frac{g'^{2}}{2A} - \frac{g^2}{r^2} - \frac{\sigma}{6}g^6 \right ];
\\
p &=& T_{22} = T_{33}=
\frac{1}{8\pi\; G}
\left [ -\frac{g'^{2}}{2A} - \frac{\sigma}{6}g^6 \right ].
\end{eqnarray}
Assuming that in the asymptotic region $A(r)\to 1$ and
$g(r)\to\frac{a}{\sqrt r}$ we get:
\begin{equation}
\rho\approx \frac{3a^2}{32 \pi\; G\; r^3}; \quad
\tau\approx -\frac{a^2}{16 \pi\; G\; r^3}; \quad
p \approx \frac{a^2}{32 \pi\; G\; r^3}.
\end{equation}
{From} the Einstein tensor for the wormhole, taking $r$ large, we
obtain the exact result $\rho = 0$ and the approximations

\begin{equation}
\tau \approx -\frac{1}{8\pi\; G}
\left [ \frac{2m \kappa}{(\lambda+\kappa)r^3} \right ];
%
\quad p \approx \frac{1}{8\pi\; G}
\left [ \frac{m\kappa}{(\lambda+\kappa)r^3}\right ].
\end{equation}
Of course, the above stress-energy does not match with the one
generated from the scalar field. To match things we add an
{\emph{exotic}} dust distribution given by:
\begin{equation}
\rho_{d} = -\frac{1}{8\pi\; G} \frac{3a^2}{4r^3};
\qquad
\tau_{d} = p_{d} = 0.
\end{equation}
Notice that this stress energy explicitly violates the energy
conditions. 
The scalar field with a sextic interaction
immersed in this dust distribution can give rise to the matter stress
energy required for our wormhole, in the large $r$ limit.

For all parameters to exactly match we would require,
\begin{equation}
\frac{a^2}{4} = \frac{m\kappa}{\lambda+\kappa}
\end{equation}
Thus, in the large $r$ region one can obtain the stresses which
generate the metric by using the above scalar field model immersed in
a dust distribution of negative energy density.

In closing this section we remind the reader that violations of the
energy conditions, though certainly present in wormhole spacetimes,
cannot be used (given our current understanding of physics) to
automatically rule out wormhole geometries. Indeed over the last few
years the catalog of physical situations in which the energy
conditions are known to be violated has been growing~\cite{cosmo99}.
There are quite reasonable looking classical systems (non-minimally
coupled scalar fields) for which all the energy conditions (including
the null energy condition) are violated; and which lead to wormhole
geometries~\cite{barcelo1,barcelo2}. In certain branches of physics,
most notably braneworld scenarios based on some form of the
Randall--Sundrum ansatz, violations of the energy conditions are now
so ubiquitous as to be completely mainstream~\cite{branes}. Attitudes
regarding the energy conditions are changing, and their violation
(even their classical violation) is no longer the anathema it has been
in the past.
\subsection{Traversability}
\label{SS:traversability}
In the asymptotically flat 3+1 dimensional spacetime
discussed in subsection (\ref{SS:general-solution}) we have three
parameters: the mass parameter m (or, in dimensionful units
${2GM}/{c^2}$) and the two wormhole parameters $\kappa$ (this gets rid
of the horizon and `makes' the geometry a wormhole) and $\lambda$.
In order to obtain values for each of these or appropriate ratios, 
we use the well--known
traversability constraints discussed by Morris and Thorne
{\cite{mt:ajp88}}.

The first constraint as mentioned in MT is related to the acceleration
felt by the traveller. Since humans are used to feeling acceleration
of order $g$ (earth gravity) we have to ensure this in the trip. The
constraint turns out to be:
\begin{equation}
\left\vert
e^{-\Phi}\frac{d(\gamma e^{\Phi})}{dl}
\right\vert  \le \frac{g}{c^2}
\approx {1\over\hbox{light year}}.
\end{equation}
For a traveller moving through the wormhole from one universe to
another we must also ensure that the tidal forces which he has to
endure should not crush him. If $v$ is the radial velocity of the
traveller (whom we assume to be of height $2\;\mathrm{m}$), the tidal
forces are related to the projections of the Riemann tensor components
along the locally Lorentz frame moving with the traveller. The
constraints as outlined by MT are given as:
\begin{widetext}
\begin{equation}
\left\vert R_{1010}\right\vert
=
\left\vert
\left(1-\frac{b}{r}\right)
\left( -\Phi''+\frac{b'r-b}{2r(r-b)}\Phi'- {\Phi'}^2 \right)
\right\vert
\le \frac{g}{c^2 \times 2\;\mathrm{m}}
\equiv \frac{1}{(10^{10}\;\mathrm{cm})^2}
\equiv \frac{1}{(10^{5}\;\mathrm{km})^2};
\end{equation}
\begin{equation}
\left\vert R_{2020}\right\vert
=
\left\vert
\frac{\gamma^2}{2r^3}\left[
\left(\frac{v}{c}\right)^2 \left(b'-\frac{b}{r}\right) + 2 (r-b)\Phi'
\right]
\right\vert
\le \frac{g}{c^2\times 2\;\mathrm{m}}
\equiv \frac{1}{(10^{10}\;\mathrm{cm})^2}
\equiv \frac{1}{(10^{5}\;\mathrm{km})^2}.
\end{equation}
\end{widetext}
The above Riemann tensor components are obtained by transforming those
in the tetrad (frame) basis attached to the Schwarzschild coordinate
system to those in a local Lorentz frame moving with a radial velocity
$v$. (For details see~\cite{visser:book}, pp 137--143.)  For our
geometry these constraints turn out to be:
\begin{equation}
\left\vert
\frac{1}{\nu^{\frac{3}{2}}} \; \gamma \;
\frac{\lambda}{\kappa \sqrt{\nu} + \lambda\sqrt{\nu -1}}
\right\vert
\le { {2GM}/{c^2}\over1\hbox{ light year}};
\end{equation}
\begin{equation}
\left\vert
\frac{1}{\nu^3} \; \frac{\lambda\sqrt{\nu -1}}{\kappa \sqrt{\nu }
+\lambda \sqrt{\nu-1}}
\right\vert
\le \left ( \frac{2GM/c^2}{10^{5}\;\mathrm{km}}\right )^2;
\end{equation}
\begin{equation}
\left\vert
\frac{\gamma^2}{2} \; \frac{1}{\nu^3} \;
\left[
\frac{\lambda \sqrt{\nu -1}}{\kappa \sqrt{\nu }
+\lambda \sqrt{\nu-1}} - \left (\frac{v}{c}\right )^2
\right]
\right\vert
\le \left ( \frac{2GM/c^2}{10^{5}\;\mathrm{km}}\right )^2.
\end{equation}
%
%
where $\nu \equiv r/(2m)$. In particular $\nu \ge 1$ (we are now using
Schwarzschild coordinates and $\nu =1$ is the throat of the wormhole).

Picking values of $\nu$ and $v$ we can calculate the acceptable range
of values for the parameters $m$, $\kappa$, and $\lambda$ which appear
in the line element. This would determine for us the {\em actual}
geometry of a macroscopic traversable wormhole without any unknown,
to--be--determined constants. As an example we choose $\nu=1$ (at the
throat of the wormhole).  Assuming $r= 2m \equiv 10^5\;\mathrm{km}$ we
find that $\kappa/\lambda \geq 10^8\;\gamma$ and $\gamma^2\beta^2<2$,
implying $\beta<\sqrt{2/3}$.  One can obtain similar bounds by
assuming other values for the throat radius.

\section{Conclusions}
\label{S:conclusions}

Let us now summarize the results obtained.  We set out with the goal
of obtaining wormholes through a certain geometric prescription. To
this end we proposed $\rho=\rho_t=0$ as the characterizing equation on
the spacetime (which also implies $R=0$, {\emph{i.e.}}, $\rho=\rho_t$,
and, equivalently, traceless stress energy).  The shape function
$b(r)$ is determined by the condition $\rho=0$. Solving the ensuing
differential equation for the $R=0$ condition we obtain the red-shift
function $\phi(r)$. The resulting line element represents a
two-parameter family of geometries which contains Lorentzian
wormholes, naked singularities, and the Schwarzschild black
hole. Using isotropic coordinates we subsequently displayed the full
structure of the solution space of the relevant equations, discussing
the domains in $\kappa$--$\lambda$ parameter space for which these
geometries arise.

The matter stress energy for the $R=0$ solutions is obtained
thereafter through a model with a triplet scalar field in a sextic
potential, superimposed upon a dust distribution of negative energy
density.  Finally the traversability constraints are written down and
analysed.

Our aim in this paper has been to provide a prescription for obtaining
wormholes. We have proposed one such prescription which is
characterized by the equation $\rho = \rho_t = 0$. This would imply
$R=0$.  One can also generalize this further and look into the
solution space of similar characterizing equations/relations which
yield wormholes and other solutions. Additionally, instead of $R=0$
one might want to obtain constant curvature wormholes which belong to
the class of spaces known as Einstein spaces.  Future work along these
lines, will, hopefully shed light on these features in greater detail.

As a punchline, we mention that the equation we propose implies the
curious fact that the Schwarzschild black hole is not as unique as it
is believed to be --- it is intimately related to a host of
traversable wormholes; solutions to the differential equations $R=0$
and $\rho =0$. We normally discard them on account of violation of the
energy conditions.  Alternatively, taking a more liberal viewpoint, we
might retain them as tentative models awaiting confirmation through
future observations.  This is particularly pertinent in view of the
recent developments in braneworld physics, where there is a marked
change in attitude towards the energy conditions.  Attitudes now tend
to be more accommodating and liberal.

\section*{Acknowledgements}
The work reported in this paper was carried out over an
extended period of time and has been benefitted by visits of ND to
the Indian Institute of Technology, Kharagpur, India, visits
of SK and SM to the Inter--University Centre for Astronomy and Astrophysics, Pune,
India and the interaction between ND and MV during GR--16 held at Durban, South
Africa in July, 2001.
The authors thank all the institutes and organisations involved 
for making these visits possible. The research of MV was supported by the
US DOE. 



\begin{thebibliography}{66}
\bibitem{mty:prl88}
M.~S.~Morris, K.~S.~Thorne and U.~Yurtsever,
``Wormholes, Time Machines, And The Weak Energy Condition'',
Phys.\ Rev.\ Lett.\  {\bf 61} (1988) 1446.

\bibitem{mt:ajp88}
M.~S.~Morris and K.~S.~Thorne,
``Wormholes In Space-Time And Their Use For Interstellar Travel:
A Tool For Teaching General Relativity'',
Am.\ J.\ Phys.\  {\bf 56} (1988) 395.

\bibitem{egy:inc}
H. Epstein, E. Glaser, and A. Yaffe,
Nuovo Cimento {\bf 36}, 1016 (1965)

\bibitem{visser:book}
M. Visser, {\em Lorentzian wormholes: from Einstein to Hawking},
AIP Press (1995).

\bibitem{geometric}
M.~Visser and D.~Hochberg,
``Geometric wormhole throats'',
gr-qc/9710001.

\bibitem{alt:prd}
D.~Hochberg,
``Lorentzian Wormholes In Higher Order Gravity Theories'',
Phys.\ Lett.\ B {\bf 251} (1990) 349.
\\
B.~Bhawal and S.~Kar,
``Lorentzian wormholes in Einstein-Gauss-Bonnet theory'',
Phys.\ Rev.\ D {\bf 46} (1992) 2464.

\bibitem{sk:prd}
S.~Kar,
``Evolving wormholes and the weak energy condition'',
Phys.\ Rev.\ D {\bf 49} (1994) 862.
\\
S.~Kar and D.~Sahdev,
``Evolving Lorentzian wormholes'',
Phys.\ Rev.\ D {\bf 53} (1996) 722
[gr-qc/9506094].

\bibitem{mv:prd}
D.~Hochberg and M.~Visser,
``The null energy condition in dynamic wormholes'',
Phys.\ Rev.\ Lett.\  {\bf 81}, 746 (1998)
[gr-qc/9802048].
\\
D.~Hochberg and M.~Visser,
``Dynamic wormholes, anti-trapped surfaces, and energy conditions'',
Phys.\ Rev.\ D {\bf 58} (1998) 044021
[gr-qc/9802046].

\bibitem{hochberg}
D.~Hochberg and M.~Visser,
``General dynamic wormholes and violation of the null energy condition'',
gr-qc/9901020.

\bibitem{ms:01}
J.~G.~Cramer, R.~L.~Forward, M.~S.~Morris, M.~Visser, G.~Benford and
G.~A.~Landis,
``Natural wormholes as gravitational lenses'',
Phys.\ Rev.\ D {\bf 51} (1995) 3117
[astro-ph/9409051].
\\
L.~A.~Anchordoqui, G.~E.~Romero, D.~F.~Torres and I.~Andruchow,
``In search for natural wormholes'',
Mod.\ Phys.\ Lett.\ A {\bf 14} (1999) 791
[astro-ph/9904399].
\\
M.~Safonova, D.~F.~Torres and G.~E.~Romero,
``Microlensing by natural wormholes: Theory and simulations'',
[gr-qc/0105070].
\\
M.~Safonova, D.~F.~Torres and G.~E.~Romero,
``Macrolensing signatures of large-scale violations of the
energy conditions'',
Mod.\ Phys.\ Lett.\ A {\bf 16} (2001) 153
[astro-ph/0104075].
\\
E.~Eiroa, G.~E.~Romero and D.~F.~Torres,
``Chromaticity effects in microlensing by wormholes'',
Mod.\ Phys.\ Lett.\ A {\bf 16} (2001) 973
[gr-qc/0104076].

\bibitem{nkd:cs}
N.~Dadhich,
``Spherically symmetric empty space and its dual in general relativity'',
Curr.\ Sci.\  {\bf 78} (2000) 1118
[gr-qc/0003018].

\bibitem{nkd:98}
N.~Dadhich,
``A duality relation: Global monopole and texture'',
gr-qc/9712021.
\\
N.~Dadhich,
``On electrogravity duality'',
Mod.\ Phys.\ Lett.\ A {\bf 14} (1999) 337
[gr-qc/9805068].

\bibitem{skds:prd}
S.~Kar and D.~Sahdev,
``Restricted class of traversable wormholes with traceless matter'',
Phys.\ Rev.\ D {\bf 52} (1995) 2030.

\bibitem{frolov}
V.~P.~Frolov,
``Vacuum polarization in a locally static multiply connected
space-time and a time machine problem'',
Phys.\ Rev.\ D {\bf 43} (1991) 3878.

\bibitem{novikov}
V.~P.~Frolov and I.~D.~Novikov,
``Physical Effects In Wormholes And Time Machine'',
Phys.\ Rev.\ D {\bf 42} (1990) 1057.

\bibitem{bv:gm}
M.~Barriola and A.~Vilenkin,
``Gravitational Field Of A Global Monopole'',
Phys.\ Rev.\ Lett.\  {\bf 63} (1989) 341.

\bibitem{cosmo99}
M.~Visser and C.~Barcelo,
``Energy conditions and their cosmological implications'',
gr-qc/0001099.

\bibitem{barcelo1}
C.~Barcelo and M.~Visser,
``Traversable wormholes from massless conformally coupled scalar fields'',
Phys.\ Lett.\ B {\bf 466} (1999) 127
[gr-qc/9908029].

\bibitem{barcelo2}
C.~Barcelo and M.~Visser,
``Scalar fields, energy conditions, and traversable wormholes'',
Class.\ Quant.\ Grav.\  {\bf 17} (2000) 3843
[gr-qc/0003025].

\bibitem{branes}
C.~Barcelo and M.~Visser,
``Brane surgery: Energy conditions, traversable wormholes, and voids'',
Nucl.\ Phys.\ B {\bf 584} (2000) 415
[hep-th/0004022].

\end{thebibliography}
\end{document}